\documentclass{iaus}
\usepackage{graphicx}
\usepackage{natbib}

\def\jnl@style#1{{\rmfamily#1}}%
\def\jref@jnl#1{{\jnl@style#1}}%

\newcommand\aj{\jref@jnl{AJ}}%
\newcommand\araa{\jref@jnl{ARA\&A}}%
\newcommand\apj{\jref@jnl{ApJ}}%
\newcommand\apjl{\jref@jnl{ApJ}}%
\newcommand\apjs{\jref@jnl{ApJS}}%
\newcommand\ao{\jref@jnl{Appl.~Opt.}}%
\newcommand\apss{\jref@jnl{Ap\&SS}}%
\newcommand\aap{\jref@jnl{A\&A}}%
\newcommand\aapr{\jref@jnl{A\&A~Rev.}}%
\newcommand\aaps{\jref@jnl{A\&AS}}%
\newcommand\azh{\jref@jnl{AZh}}%
\newcommand\baas{\jref@jnl{BAAS}}%
\newcommand\jrasc{\jref@jnl{JRASC}}%
\newcommand\memras{\jref@jnl{MmRAS}}%
\newcommand\mnras{\jref@jnl{MNRAS}}%
\newcommand\pra{\jref@jnl{Phys.~Rev.~A}}%
\newcommand\prb{\jref@jnl{Phys.~Rev.~B}}%
\newcommand\prc{\jref@jnl{Phys.~Rev.~C}}%
\newcommand\prd{\jref@jnl{Phys.~Rev.~D}}%
\newcommand\pre{\jref@jnl{Phys.~Rev.~E}}%
\newcommand\prl{\jref@jnl{Phys.~Rev.~Lett.}}%
\newcommand\pasp{\jref@jnl{PASP}}%
\newcommand\pasj{\jref@jnl{PASJ}}%
\newcommand\qjras{\jref@jnl{QJRAS}}%
\newcommand\skytel{\jref@jnl{S\&T}}%
\newcommand\solphys{\jref@jnl{Sol.~Phys.}}%
\newcommand\sovast{\jref@jnl{Soviet~Ast.}}%
\newcommand\ssr{\jref@jnl{Space~Sci.~Rev.}}%
\newcommand\zap{\jref@jnl{ZAp}}%
\newcommand\nat{\jref@jnl{Nature}}%
\newcommand\iaucirc{\jref@jnl{IAU~Circ.}}%
\newcommand\aplett{\jref@jnl{Astrophys.~Lett.}}%
\newcommand\apspr{\jref@jnl{Astrophys.~Space~Phys.~Res.}}%
\newcommand\bain{\jref@jnl{Bull.~Astron.~Inst.~Netherlands}}%
\newcommand\fcp{\jref@jnl{Fund.~Cosmic~Phys.}}%
\newcommand\gca{\jref@jnl{Geochim.~Cosmochim.~Acta}}%
\newcommand\grl{\jref@jnl{Geophys.~Res.~Lett.}}%
\newcommand\jcp{\jref@jnl{J.~Chem.~Phys.}}%
\newcommand\jgr{\jref@jnl{J.~Geophys.~Res.}}%
\newcommand\jqsrt{\jref@jnl{J.~Quant.~Spec.~Radiat.~Transf.}}%
\newcommand\memsai{\jref@jnl{Mem.~Soc.~Astron.~Italiana}}%
\newcommand\nphysa{\jref@jnl{Nucl.~Phys.~A}}%
\newcommand\physrep{\jref@jnl{Phys.~Rep.}}%
\newcommand\physscr{\jref@jnl{Phys.~Scr}}%
\newcommand\planss{\jref@jnl{Planet.~Space~Sci.}}%
\newcommand\procspie{\jref@jnl{Proc.~SPIE}}%

\title[Sink particles in AMR and SPH] 
{Implementing and comparing sink particles in AMR and SPH}

\author[Federrath et al.]   
{Christoph~Federrath$^{1}$, Robi~Banerjee$^{1}$, Daniel~Seifried$^{1}$, Paul~C.~Clark$^{1}$, \and Ralf~S.~Klessen$^{1}$}

\affiliation{$^1$Zentrum f\"ur Astronomie der Universit\"at Heidelberg, \\Institut f\"ur Theoretische Astrophysik, Albert-Ueberle-Str.~2, 69120 Heidelberg, Germany \\email: {\tt chfeder@ita.uni-heidelberg.de}}

\pubyear{2010}
\volume{270}  
\pagerange{xxx--xxx}
\jname{Computational Star Formation}
\editors{eds. Alves, Elmegreen, Girart, Trimble}
\begin{document}

\maketitle

\begin{abstract}
We implemented sink particles in the Adaptive Mesh Refinement (AMR) code FLASH to model the gravitational collapse and accretion in turbulent molecular clouds and cores. Sink particles are frequently used to measure properties of star formation in numerical simulations, such as the star formation rate and efficiency, and the mass distribution of stars. We show that only using a density threshold for sink particle creation is insufficient in case of supersonic flows, because the density can exceed the threshold in strong shocks that do not necessarily lead to local collapse. Additional physical collapse indicators have to be considered. We apply our AMR sink particle module to the formation of a star cluster, and compare it to a Smoothed Particle Hydrodynamics (SPH) code with sink particles. Our comparison shows encouraging agreement of gas and sink particle properties between the AMR and SPH code.
\end{abstract}

\firstsection 

\section{Introduction}
Stars form in turbulent, magnetized molecular clouds by local gravitational collapse of dense gas cores \citep{MacLowKlessen2004}. Modeling this process in computer simulations is extremely difficult. It is necessary to follow the freefall collapse of each individual star, while keeping track of the global evolution of the entire cloud at the same time. The fundamental numerical difficulty is that the freefall timescale decreases with increasing gas density: $t_\mathrm{ff} = \sqrt{3\pi/32 G\rho}$. Modeling each individual collapse and following the large-scale evolution of the cloud over several global freefall times in a single magnetohydrodynamical calculation is beyond the capabilities of modern numerical schemes and supercomputers. Thus, if one wants to model the evolution of such a cloud, the individual runaway collapse must be cut-off in a controlled way and replaced by a subgrid model.

There are two such subgrid models: heating the gas or using sink particles. In the first approach the gas is heated up above a given density threshold to prevent artificial fragmentation beyond the resolution limit for collapse \citep{TrueloveEtAl1997}. There are two problems with the heating approach. First, the Courant timestep decreases, because the sound speed increases, and second, the gas equation of state is changed above the density threshold. For molecular clouds, heating of the gas may only occur for densities $\rho\gtrsim10^{-14}\,\mathrm{g}\,\mathrm{cm}^{-3}$ in the optically thick regime \citep{Larson1969,Penston1969}. However, gas can become denser than the threshold value in shocks that do not necessarily lead to the formation of a gravitationally bound structure. Thus, shocked gas not going into freefall collapse will be heated up artificially at least in the case where the threshold density for heating is in the isothermal regime.

The alternative subgrid model is to use Lagrangian sink particles, a method invented by \citet*{BateBonnellPrice1995} for Smoothed Particle Hydrodynamics (SPH), and first adopted for Eulerian, Adaptive Mesh Refinement (AMR) by \citet*{KrumholzMcKeeKlein2004}. If the gas reaches a given density, a Lagrangian, accreting sink particle is introduced. However, sink particles are supposed to represent bound objects, and thus, a density threshold for their creation is insufficient. Compression in shocks can temporarily create densities larger than the threshold without triggering gravitational collapse. Previous grid-based implementations of sink particles are mostly based on a density threshold criterion. Here, we present an implementation of sink particles for the Eulerian, AMR code FLASH \citep{FryxellEtAl2000} that uses a series of checks, such that only bound and collapsing gas is turned into sink particles. We show that the star formation efficiency and the number of fragments is overestimated, if additional, physical checks in addition to the density threshold are ignored in the isothermal regime.

The main results presented here were previously published in \citet{FederrathBanerjeeClarkKlessen2010}. However, we discuss two new tests for linear and angular momentum conservation in \S\ref{sec:mom}. In \S\ref{sec:checkscomp} we present the physical checks necessary to avoid spurious sink particle creation and present the main results of the AMR--SPH comparison of sink particles.

\section{Sink particle creation checks and AMR--SPH comparison} \label{sec:checkscomp}

\begin{figure}[tb]
\begin{center}
\def\arraystretch{1.0}
\begin{tabular}{c}
\includegraphics[height=0.42\linewidth]{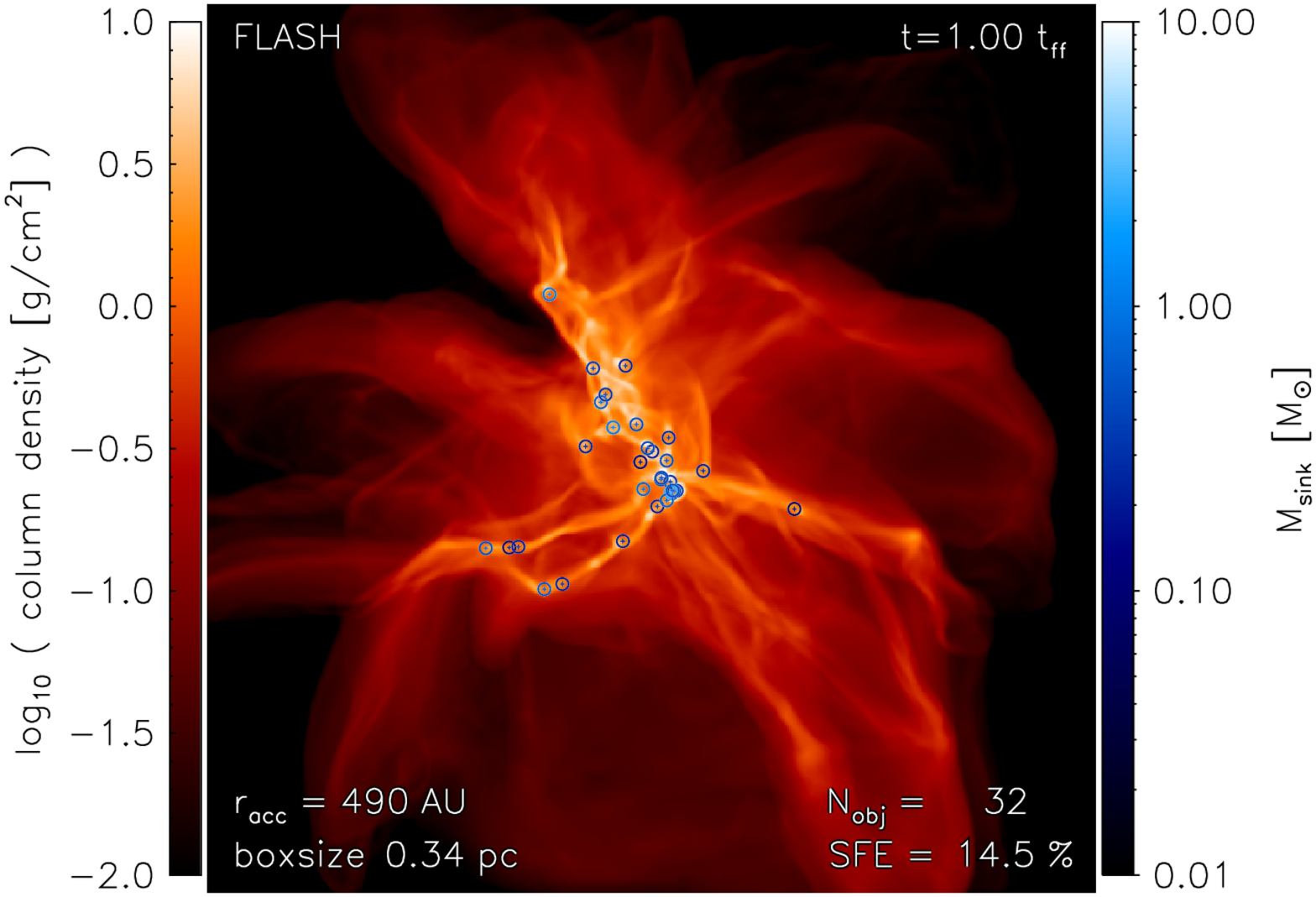} \\
\includegraphics[height=0.42\linewidth]{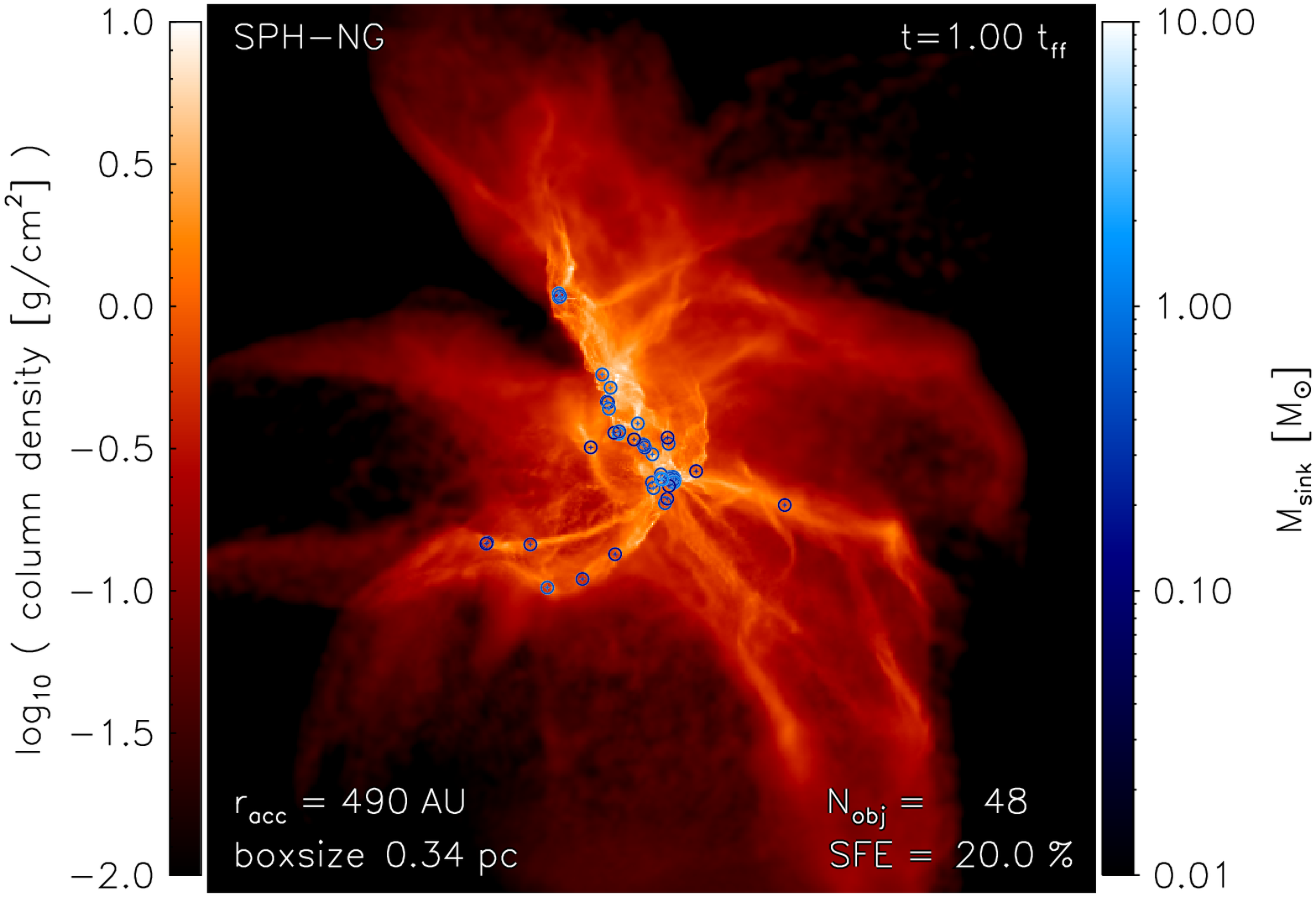} \\
\includegraphics[height=0.42\linewidth]{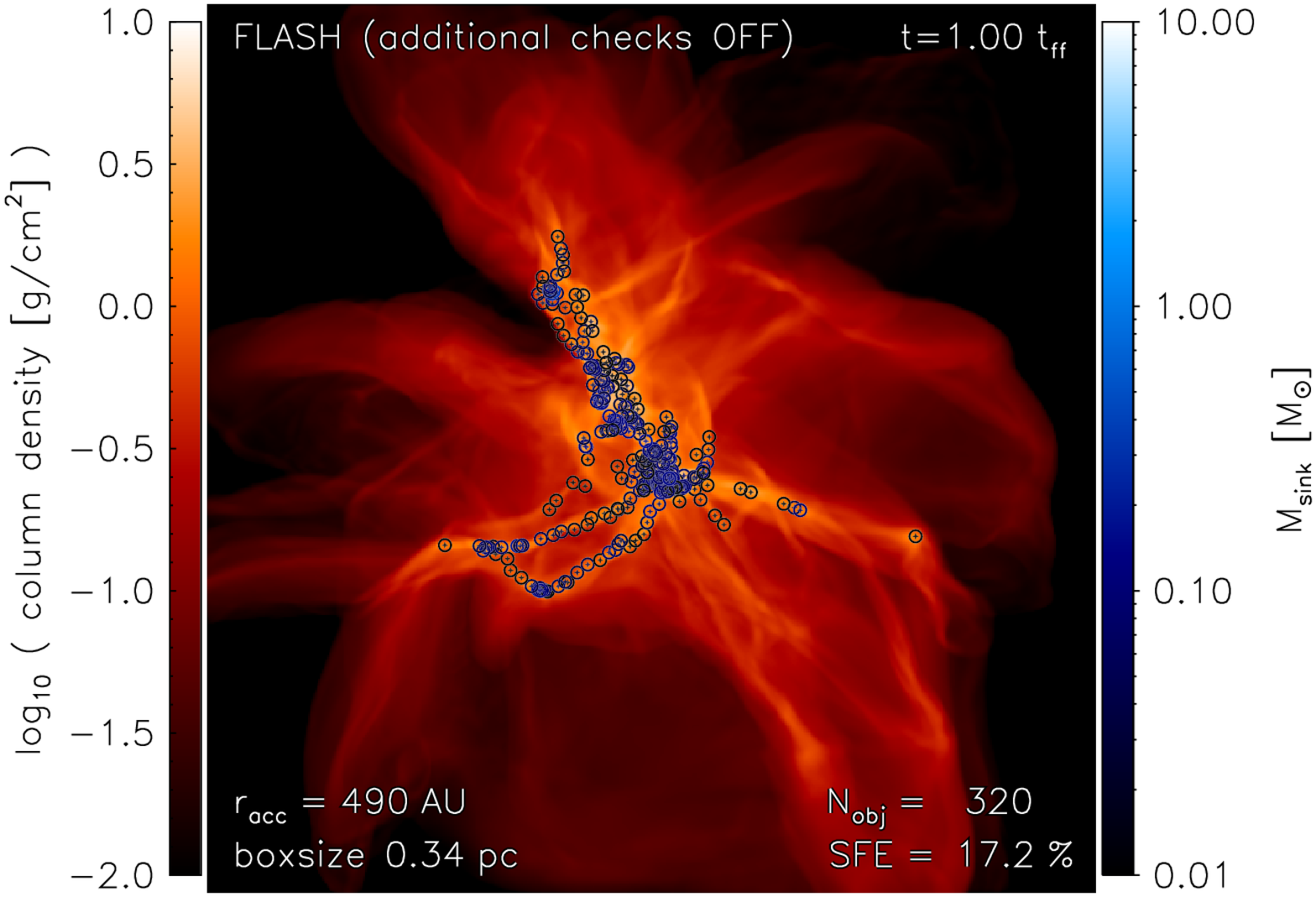}
\end{tabular}
\end{center}
\caption{The formation of a stellar cluster from a $100\,\mathrm{M}_\odot$ turbulent cloud after one global freefall time. \emph{Top:} FLASH (default settings), \emph{middle:} SPH-NG, and \emph{bottom:} FLASH (first four checks switched off). See \citet{FederrathBanerjeeClarkKlessen2010} for further details.}
\label{fig:checkscomp}
\end{figure}

We refer the reader to \citet[][\S2.2]{FederrathBanerjeeClarkKlessen2010} for a detailed discussion and for the implementation of the sink particle creation checks. In summary, for successful sink creation the gas exceeding the density threshold must also
\begin{enumerate}
\item be converging (along each cardinal direction individually),
\item have a central gravitational potential minimum,
\item be Jeans-unstable (including magnetic pressure),
\item be bound (including magnetic energy),
\item be on the highest level of the AMR (i.e., Jeans length resolved),
\item not be within accretion radius of existing sinks (then accretion checks apply).
\end{enumerate}
During testing it turned out that the first two checks are particularly important to avoid spurious sink particle creation, however, the relative importance of each of the individual checks should be investigated in more detail (Wadsley et al., in prep.).

Fig.~\ref{fig:checkscomp} shows the comparison of FLASH (all checks on; top panel) against FLASH (first four checks switched off; bottom panel), clearly demonstrating the importance of the physical sink creation checks in the isothermal regime. The middle panel of Fig.~\ref{fig:checkscomp} shows the SPH-NG run, exhibiting some differences to the FLASH run, which can be attributed to the slightly faster collapse of the cloud core in the SPH run. This is most likely a consequence of the faster decay of the supporting initial turbulence due to the slightly higher viscosity in SPH \citep{PriceFederrath2010}. After correcting for this and comparing at times when 26\% of the gas has been accreted onto sinks, the FLASH and SPH-NG runs are in very good agreement with 49 and 50 sink particles formed, respectively, and having similar mass distributions \citep[see \S4 in][]{FederrathBanerjeeClarkKlessen2010}.

\section{Momentum and angular momentum conservation test} \label{sec:mom}

\begin{figure}[tb]
\begin{center}
  \def\arraystretch{1.0}
  \begin{tabular}{cc}
    \begin{tabular}{cc}
      \includegraphics[width=0.26\linewidth]{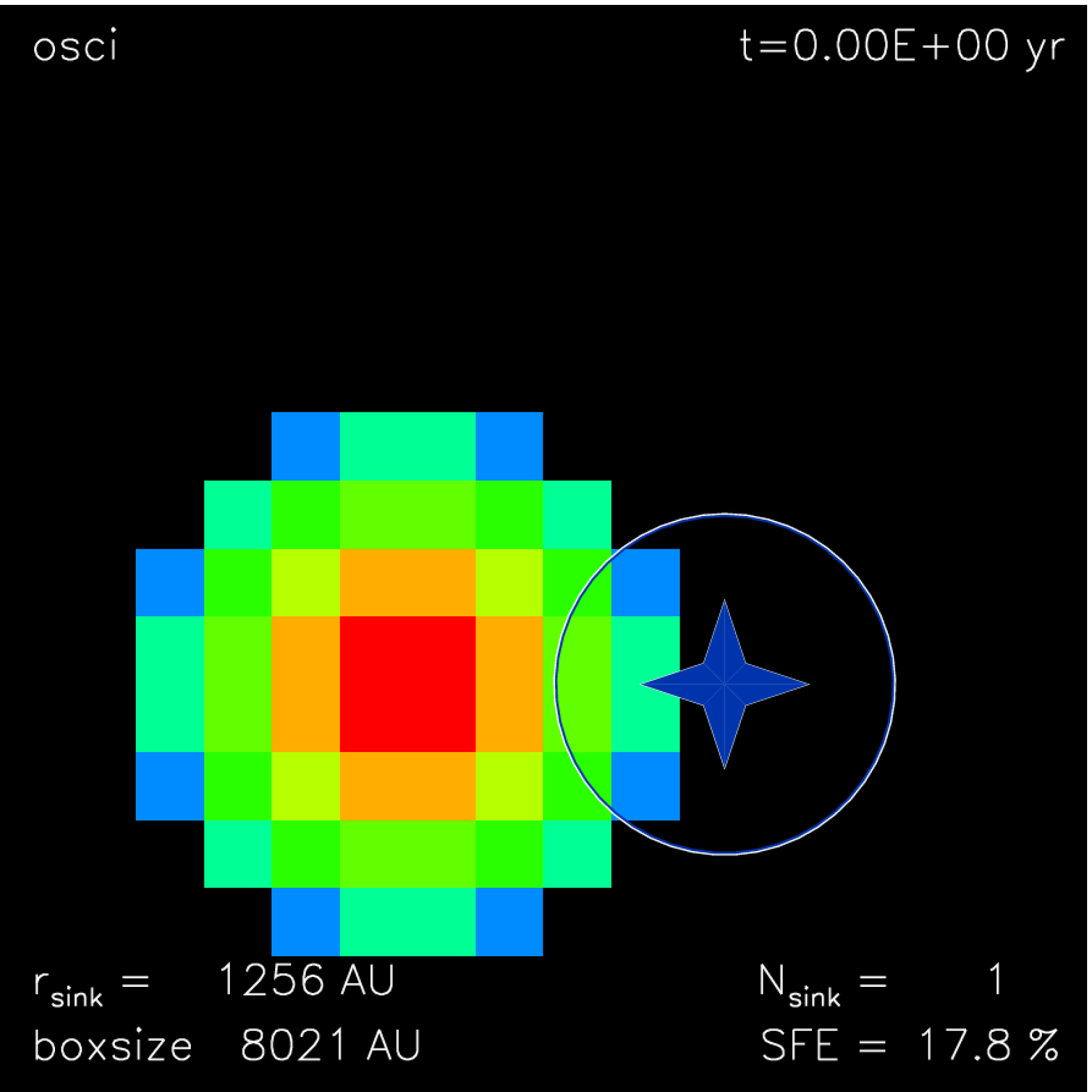} &
      \includegraphics[width=0.26\linewidth]{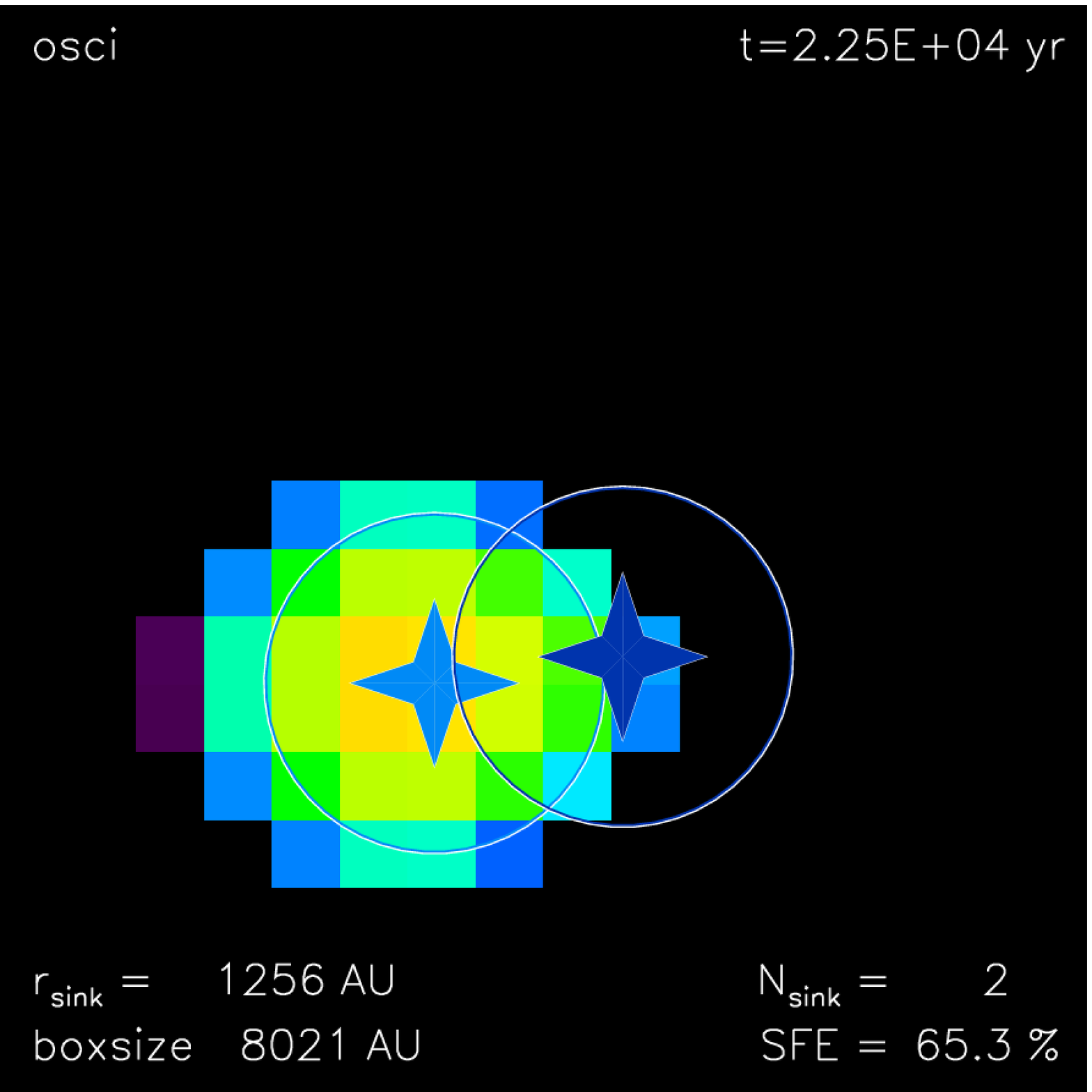} \\
      \includegraphics[width=0.26\linewidth]{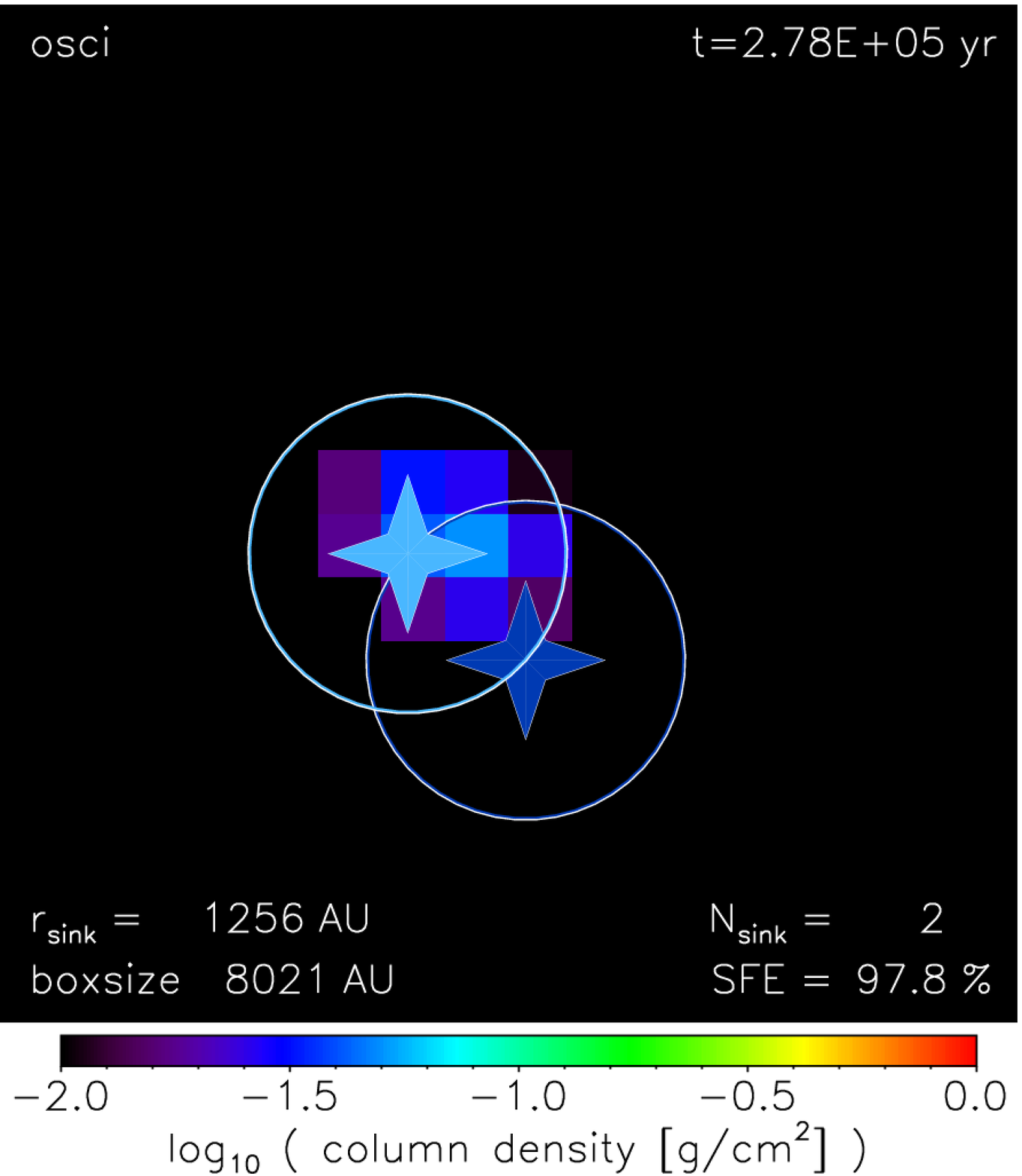} &
      \includegraphics[width=0.26\linewidth]{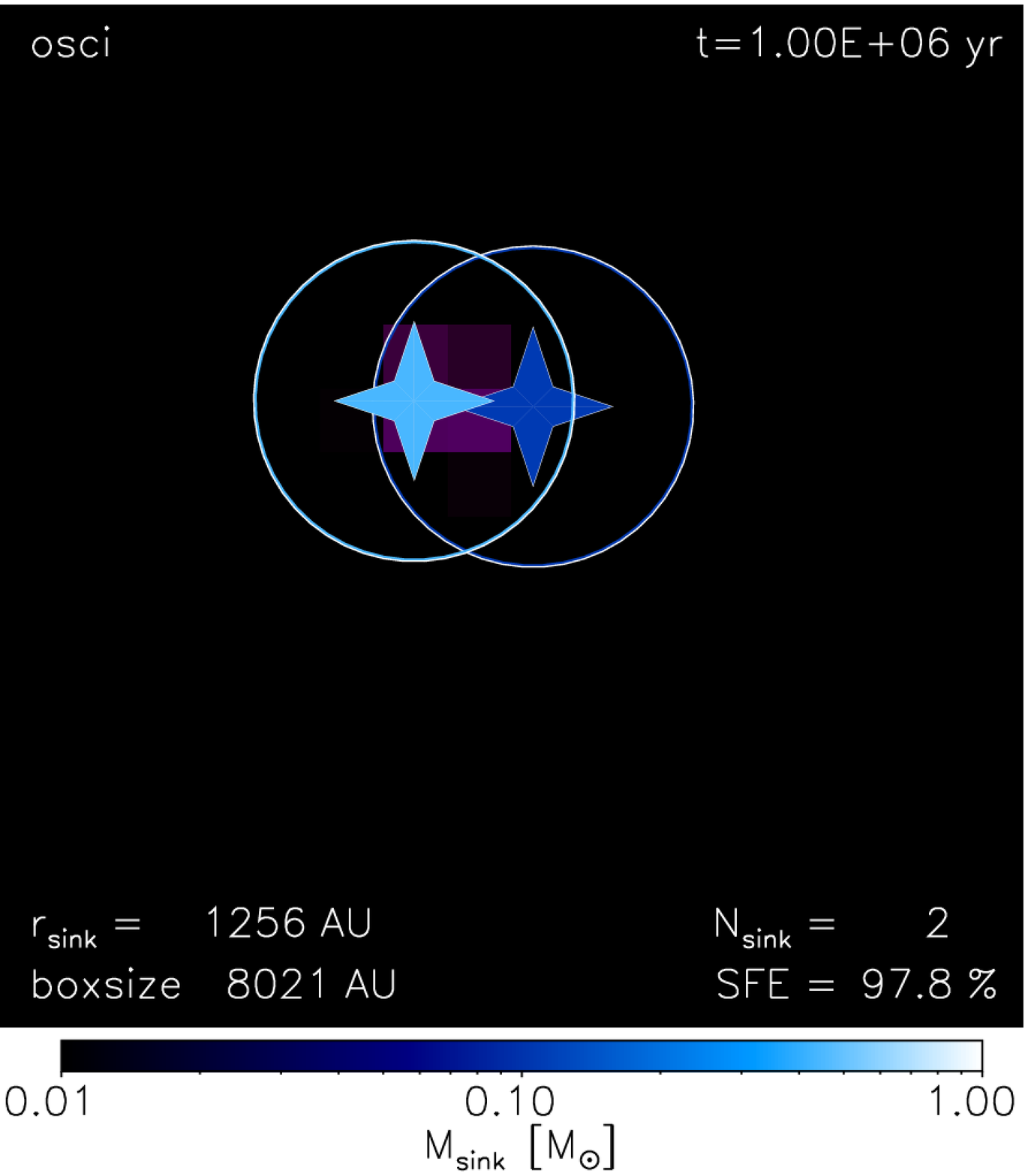}
    \end{tabular}
    &
    \begin{tabular}{c}
      \includegraphics[width=0.4\linewidth]{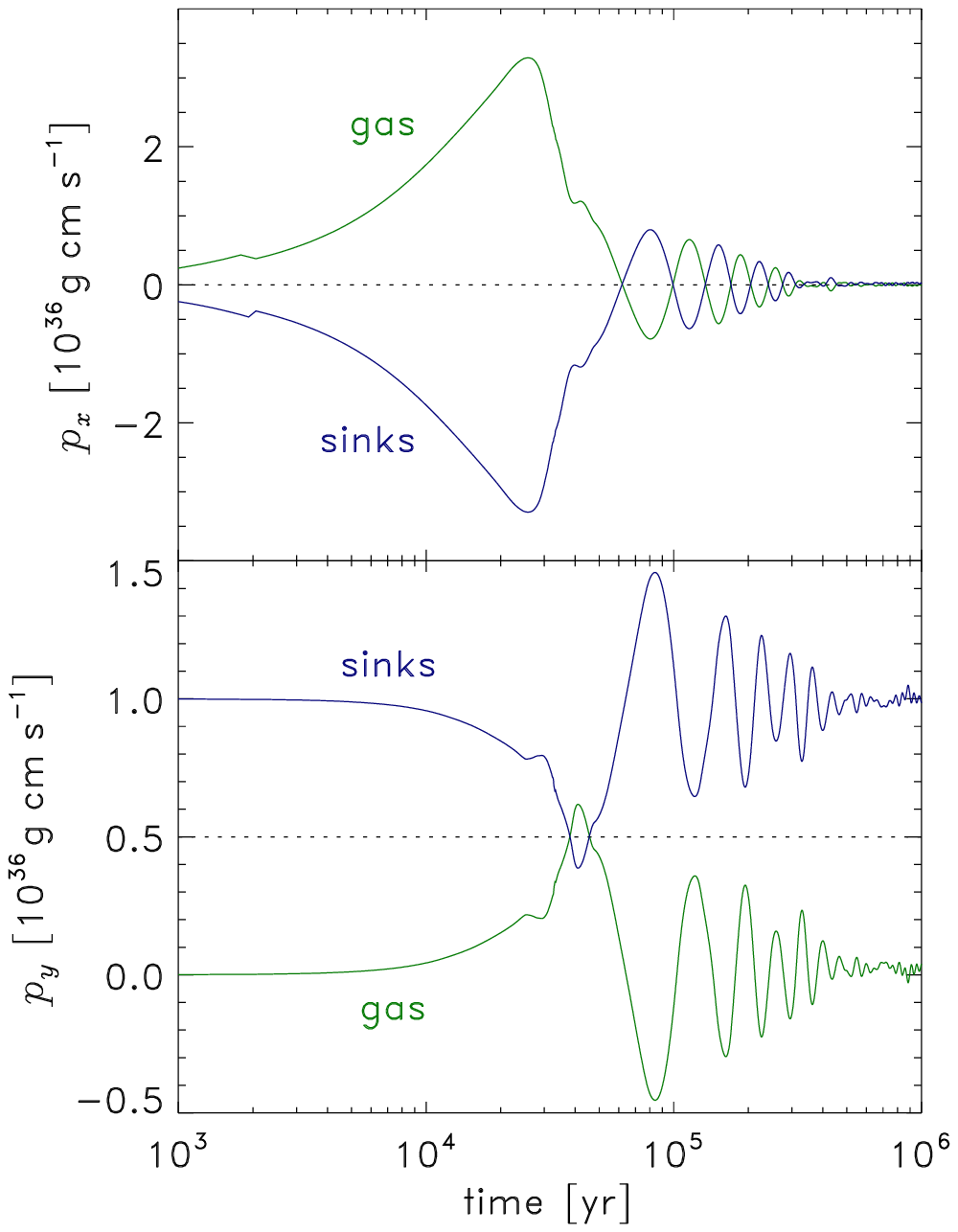}
    \end{tabular}
  \end{tabular}
\end{center}
\caption{\emph{Left panels:} time evolution of the column density in the momentum conservation test. \emph{Right panels:} time evolution of the momenta of sink particles and gas in $x$- (\emph{top}) and $y$-direction (\emph{bottom}). For momentum conservation to hold, the momenta of sinks and gas must be symmetric about the dotted line, which indicates half the initial total momentum.}
\label{fig:osci}
\end{figure}

In \citet{FederrathBanerjeeClarkKlessen2010} we performed a suite of test simulations for the new sink particle module in FLASH, including circular and highly eccentric orbits, the collapse of a Bonnor-Ebert sphere and a singular isothermal sphere, and a rotating cloud core fragmentation test. Here we add a momentum conservation test shown in Fig.~\ref{fig:osci}, where we initialized a $0.46\,\mathrm{M}_\odot$ core (with very low resolution for testing purposes) at rest and a $0.1\,\mathrm{M}_\odot$ sink particle with an initial momentum of $p_y=10^{36}\,\mathrm{g}\,\mathrm{cm}\,\mathrm{s}^{-1}$ in $y$-direction. Both gas core and sink particle are offset from the center. Fig.~\ref{fig:osci} (left panels) show column density snapshots of the time evolution of the system. The initial gas core collapses and a second sink particle forms and accretes almost all the initial gas mass, while the initial sink particle only has time to accrete about 2\% of the gas. The two sink particles are then followed for 20 orbits up to $t_\mathrm{end}=10^6\,\mathrm{yr}$. The right panels of Fig.~\ref{fig:osci} show the time evolution of the momentum. Momentum is conserved to within 3\% over the whole duration even in this very low resolution run, shown by the symmetry of sink and gas momenta. Kinks in the momentum evolution indicate strong accretion events.

\begin{figure}[tb]
\begin{center}
  \def\arraystretch{1.0}
  \begin{tabular}{cc}
      \includegraphics[width=0.48\linewidth]{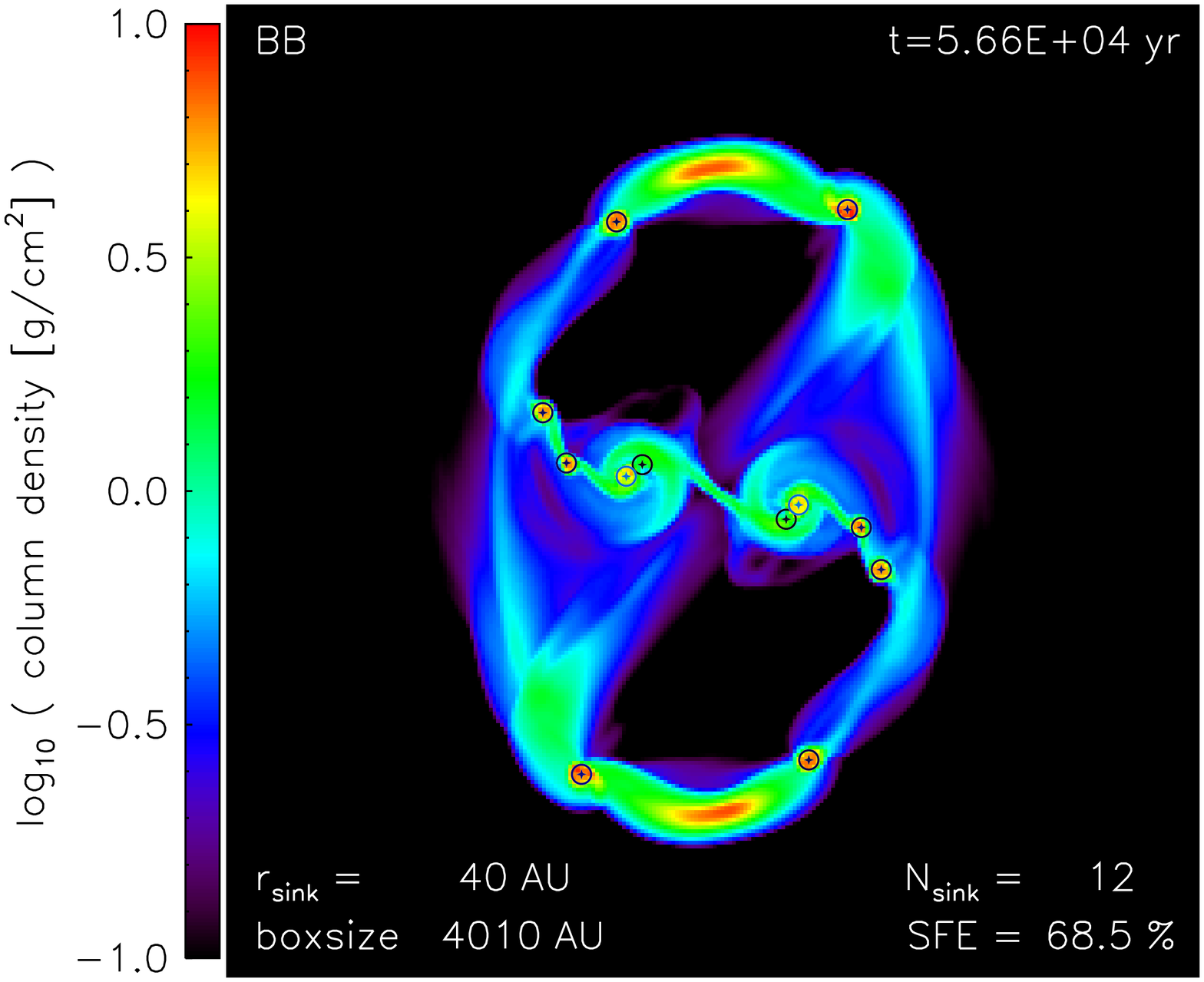} &
      \includegraphics[width=0.48\linewidth]{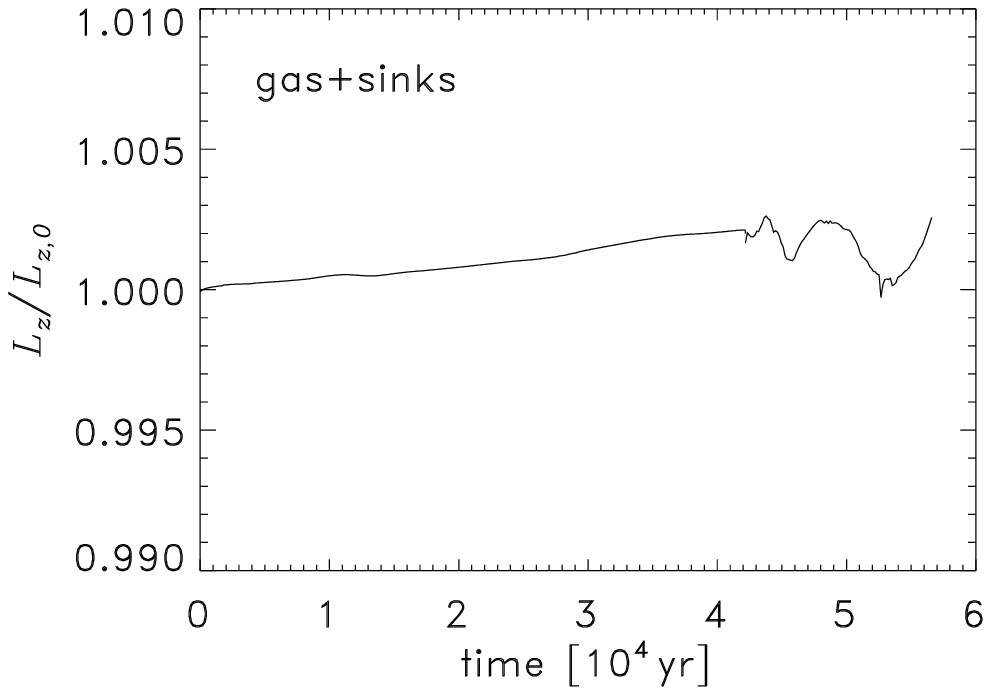}
  \end{tabular}
\end{center}
\caption{\emph{Left:} Column density snapshot of a variant of the \citet{BossBodenheimer1979} test, followed to very late times when 12 fragments have formed and 68.5\% of the gas has been accreted by sinks. \emph{Right:} Shows that angular momentum is conserved to within 0.5\% for all times.}
\label{fig:bb}
\end{figure}

For testing angular momentum conservation, we use a high-resolution model of the \citet{BossBodenheimer1979} test. Figure~\ref{fig:bb} (left) shows the column density of the disk (face on in the $x$-$y$-plane) for the last time frame computed, where 12 fragments have formed in the disk, which turned into a ring-like structure at late times. The right panel shows the time evolution of the total angular momentum, $L_z$. Angular momentum is conserved to within 0.5\% \citep[see][for a comparison of angular momentum conservation in SPH and AMR]{CommerconEtAl2008}.

\section{Conclusions} \label{sec:conclusions}
We introduced and tested a new sink particle method for the AMR code FLASH in \citet{FederrathBanerjeeClarkKlessen2010}. In addition to the tests shown there we presented a linear and angular momentum conservation test in \S\ref{sec:mom}. A comparison of gas and sink particle properties showed encouraging agreement with the SPH-NG code \citep{BateBonnellPrice1995}. More recently, another SPH code, GASOLINE \citep{WadsleyStadelQuinn2004}, also showed very good agreement, if the sink particle creation checks outlined in \S\ref{sec:checkscomp} are used (Wadsley et al., in prep.).

\begin{figure}[tb]
 \centerline{\includegraphics[width=0.7\linewidth]{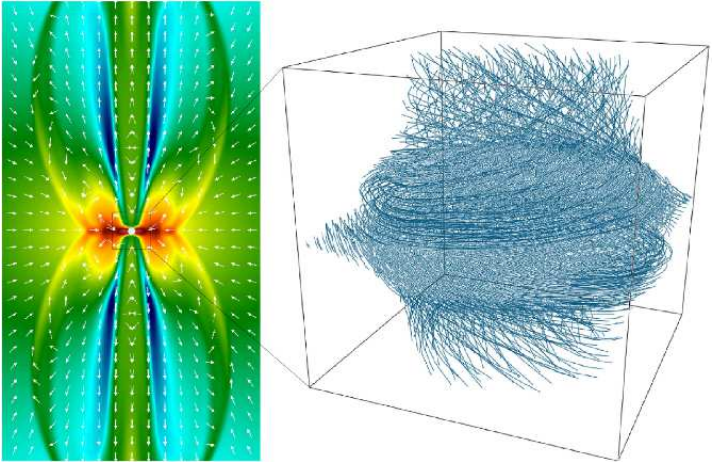}}
 \caption{\emph{Left:} Density slice through a rotating, magnetized disk. A bipolar outflow (velocity vectors are overlaid on the density) has formed and propagated roughly $5,000\,\mathrm{AU}$ from the disk midplane. The central sink particle has a radius of $80\,\mathrm{AU}$. \emph{Right:} Shows the three-dimensional magnetic field structure in the inner $800\,\mathrm{AU}$ of the disk. The magnetic field is wound up strongly and launches the outflows shown in the left image.}
 \label{fig:outflow}
 \end{figure}

Some open issues remain concerning the modeling of magnetic fields in combination with sink particles in SPH \citep{PriceBate2008,Price2010}. The problem is that SPH particles are accreted inside the sink particle radius and are thus lost as resolution elements for the magnetic field. The advantage of using a grid-based implementation of sink particles is that the stencil for the magnetic field remains when gas is accreted. The geometry of the magnetic field thus remains intact. We also performed simulations of the collapse of magnetized, rotating cloud cores, which self-consistently produce bipolar outflows (Duffin et al.; Seifried et al., in prep.), showing that our sink particle approach works in combination with magnetic fields (Figure~\ref{fig:outflow}).

\bibliographystyle{apj}

\begin{thebibliography}{14}
\expandafter\ifx\csname natexlab\endcsname\relax\def\natexlab#1{#1}\fi

\bibitem[{{Bate} {et~al.}(1995){Bate}, {Bonnell}, \&
  {Price}}]{BateBonnellPrice1995}
{Bate}, M.~R., {Bonnell}, I.~A., \& {Price}, N.~M. 1995, \mnras, 277, 362

\bibitem[{{Boss} \& {Bodenheimer}(1979)}]{BossBodenheimer1979}
{Boss}, A.~P., \& {Bodenheimer}, P. 1979, \apj, 234, 289

\bibitem[{{Commer{\c c}on} {et~al.}(2008){Commer{\c c}on}, {Hennebelle},
  {Audit}, {Chabrier}, \& {Teyssier}}]{CommerconEtAl2008}
{Commer{\c c}on}, B., {Hennebelle}, P., {Audit}, E., {Chabrier}, G., \&
  {Teyssier}, R. 2008, \aap, 482, 371

\bibitem[{{Federrath} {et~al.}(2010){Federrath}, {Banerjee}, {Clark}, \&
  {Klessen}}]{FederrathBanerjeeClarkKlessen2010}
{Federrath}, C., {Banerjee}, R., {Clark}, P.~C., \& {Klessen}, R.~S. 2010,
  \apj, 713, 269

\bibitem[{{Fryxell} {et~al.}(2000){Fryxell}, {Olson}, {Ricker}, {Timmes},
  {Zingale}, {Lamb}, {MacNeice}, {Rosner}, {Truran}, \&
  {Tufo}}]{FryxellEtAl2000}
{Fryxell}, B., {Olson}, K., {Ricker}, P., {Timmes}, F.~X., {Zingale}, M.,
  {Lamb}, D.~Q., {MacNeice}, P., {Rosner}, R., {Truran}, J.~W., \& {Tufo}, H.
  2000, \apjs, 131, 273

\bibitem[{{Krumholz} {et~al.}(2004){Krumholz}, {McKee}, \&
  {Klein}}]{KrumholzMcKeeKlein2004}
{Krumholz}, M.~R., {McKee}, C.~F., \& {Klein}, R.~I. 2004, \apj, 611, 399

\bibitem[{{Larson}(1969)}]{Larson1969}
{Larson}, R.~B. 1969, \mnras, 145, 271

\bibitem[{{Mac Low} \& {Klessen}(2004)}]{MacLowKlessen2004}
{Mac Low}, M.-M., \& {Klessen}, R.~S. 2004, Reviews of Modern Physics, 76, 125

\bibitem[{{Penston}(1969)}]{Penston1969}
{Penston}, M.~V. 1969, \mnras, 144, 425

\bibitem[{{Price}(2010)}]{Price2010}
{Price}, D.~J. 2010, \mnras, 401, 1475

\bibitem[{{Price} \& {Bate}(2008)}]{PriceBate2008}
{Price}, D.~J., \& {Bate}, M.~R. 2008, \mnras, 385, 1820

\bibitem[{{Price} \& {Federrath}(2010)}]{PriceFederrath2010}
{Price}, D.~J., \& {Federrath}, C. 2010, \mnras, 406, 1659

\bibitem[{{Truelove} {et~al.}(1997){Truelove}, {Klein}, {McKee}, {Holliman},
  {Howell}, \& {Greenough}}]{TrueloveEtAl1997}
{Truelove}, J.~K., {Klein}, R.~I., {McKee}, C.~F., {Holliman}, II, J.~H.,
  {Howell}, L.~H., \& {Greenough}, J.~A. 1997, \apjl, 489, L179

\bibitem[{{Wadsley} {et~al.}(2004){Wadsley}, {Stadel}, \&
  {Quinn}}]{WadsleyStadelQuinn2004}
{Wadsley}, J.~W., {Stadel}, J., \& {Quinn}, T. 2004, New Astronomy, 9, 137

\end{thebibliography}

\end{document}